\pdfoutput=1

\documentclass[conference]{IEEEtran}
\renewcommand\IEEEkeywordsname{Keywords}
\usepackage{hyperref}
\usepackage[pdftex]{graphicx}
\usepackage{amsmath}
\usepackage{eqparbox}
\usepackage[labelformat=empty]{caption}
\usepackage{hyperref}
\usepackage{tabularx,booktabs}
\usepackage{multirow}
\usepackage{silence}
\usepackage{floatrow}
\usepackage{cite}
\hypersetup{%
  pdfborder = {0 0 0},
  bookmarksdepth = subsubsection,
  citecolor = DarkGreen,
  linkcolor = DarkBlue,
  urlcolor = DarkGreen,
  bookmarks=true
}%

\begin{document}

\title{\LARGE Advaita: Bug Duplicity Detection System}

\author{\IEEEauthorblockN{Amit Kumar,
Manohar Madanu, Hari Prakash, Lalitha Jonnavithula,
Srinivasa Rao Aravilli}
\IEEEauthorblockA{
    @Cisco Systems, Inc}
}


\maketitle
\thispagestyle{plain}
\pagestyle{plain}


\begin{abstract}
Bugs are prevalent in software development. To improve software quality, bugs are filed using a bug tracking system. Properties of a reported bug would consist of a headline, description, project, product, component that is affected by the bug and the severity of the bug. Duplicate bugs rate (\% of duplicate bugs) are in the range from single digit (1 to 9\%) to double digits (40\%) based on the product maturity , size of the code and number of engineers working on the project. Duplicate bugs range are between 9\% to 39\% in some of the open source projects like Eclipse, Firefox etc. Detection of duplicity deals with identifying whether any two bugs convey the same meaning. This detection of duplicates helps in de-duplication. Detecting duplicate bugs help reduce triaging efforts and saves time for developers in fixing the issues. Traditional natural language processing techniques are less accurate in identifying similarity between sentences. Using the bug data present in a bug tracking system, various approaches were explored including several machine learning algorithms, to obtain a viable approach that can identify duplicate bugs, given a pair of sentences(i.e. the respective bug descriptions). This approach considers multiple sets of features viz. basic text statistical features, semantic features and contextual features. These features are extracted from the headline, description and component and are subsequently used to train a classification algorithm.
\end{abstract}



\section{Introduction}
\vspace{1em}
\textbf{B}ug tracking system is one of the best practices in the maintenance of computer software. Bug reports are submitted by both engineers and users. Allowing users to submit bug reports provides engineers with continuous feedback about the operational behavior of the product. However, submitted reports may vary in quality and it is often difficult to understand what is being reported. Bug reports are initially classified through a process called triaging. An analyst, commonly called a triager, must determine whether a newly submitted bug describes a previously unreported issue or a previously reported issue. Former is referred to as an original bug and the latter as a duplicate.
\newline \par 
Manual triaging of bug reports is both challenging and time consuming. The very nature of the English language entails that two people can use vastly different language expressions to describe the same issue.\newline \par 

In this paper, a system is presented that can classify each bug reported as either an original or a duplicate by extracting a novel set of features associated with the bug report and the extent of similarity to prior reports. These features are designed to provide multiple viewpoints into the content of bug reports and offer an extensive yet diverse set of features viewpoints allowing the automated system to differentiate between original and duplicate reports. \newline \par
 The major contributions of this paper are described as follows:\newline
 \begin{itemize}
    \item Detailed description of a novel automated framework for classifying bug reports as originals or duplicates.
    \item Novel set of features associated with similarity scores, semantic, syntactical and contextual features that can be used to automatically classify bug reports.
    \item A demonstration, through empirical study, that this approach reaches high accuracy when applied to any standard project’s bug tracking software.
\end{itemize}

\begin{figure*}[ht!] 
\centering
\includegraphics[width=6.5in]{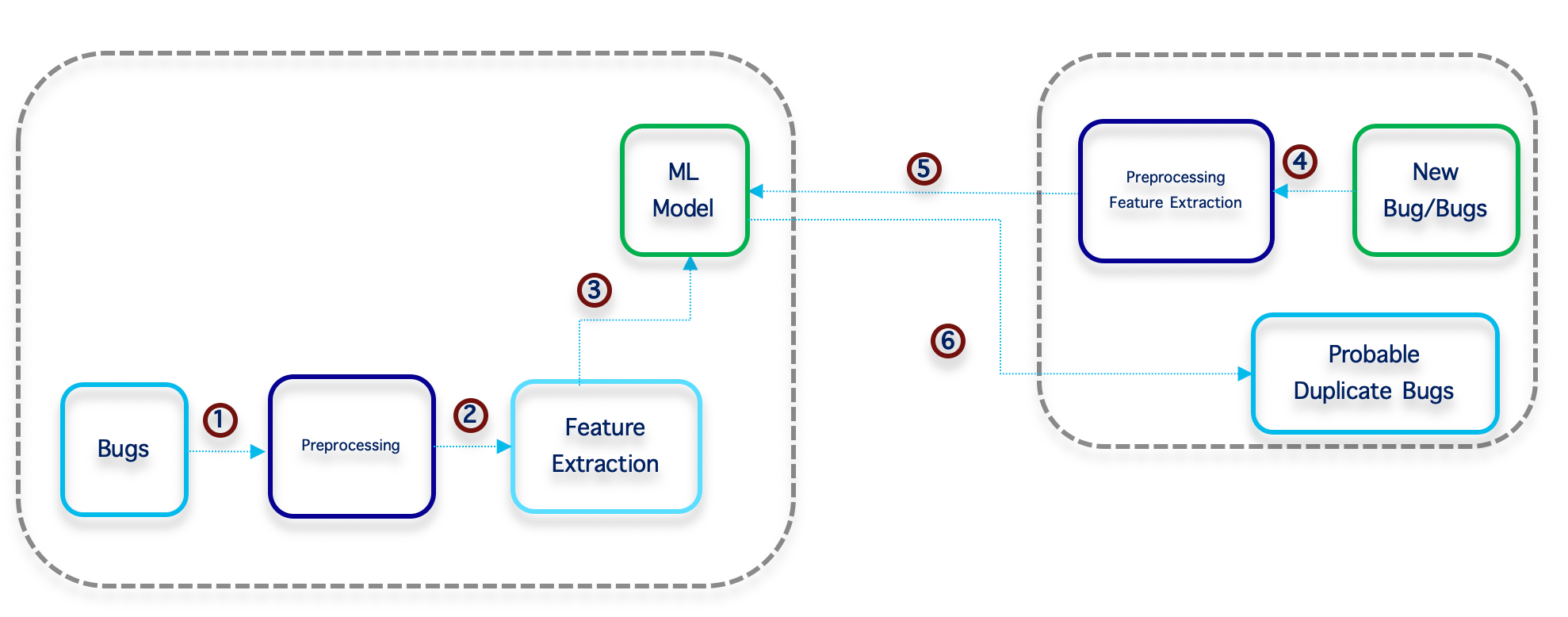}
\caption{Fig 1. The figure shows Advaita flow diagram}
\label{picture}
\end{figure*}

\vspace{1em}
\section{Related Work}
\vspace{1em}
There are several approaches that apply machine learning (ML) and/or information retrieval (IR) to (semi-)automate the process of bug de-duplication. The first approach along that
line is from Cubranic and Murphy \cite{one}. From the titles and descriptions of bug reports, keywords and developers’ ID's are extracted and used to build a text classifier using Naive Bayes technique. The classifier will recommend potential fixers based on the classification of a new report. Their prediction accuracy is around 30\% on an Eclipse bug report dataset from Jan 2002 to Sep 2002. 
Anvik et al. \cite{two} also follow a similar ML approach and improve Cubranic et al.’s work
by filtering out invalid data such as unfixed bug reports, no-longer-working or inactive developers. With 3 different classifiers (SVM, Naive Bayes, and C4.5) a precision of up to 64\% was achieved.\newline \par
Another related approach is from Bhattacharya and Neamtiu \cite{three}. Like Bugzie \cite{four}, their model is capable of incremental learning. However, in contrast to fuzzy set approach in Bugzie, they use a ML approach with Naive Bayes and Bayesian network to build classifiers for keywords extracted from reports. Therefore, Bugzie has better time efficiency and a more natural ranking scheme than ML classifiers.\newline \par
Lin et al. \cite{five} uses a ML approach with SVM and C4.5 classifiers on both textual data and non-text fields (e.g. bug type, priority, submitter, phase and module IDs). Executing on a proprietary project with 2,576 bug records, their models achieve an accuracy of up to 77.64\%. The accuracy is 63\% if module IDs were not considered. Bugzie has higher accuracy and can integrate non-text fields for further improvement. Other researchers have studied bug reports based on different types, quality and severity \cite{six, seven, eight}.\newline


\section{Advaita For Bug Duplicity Detection System}
\vspace{1em}
In order to identify duplicate bugs Advaita cleans the input data and generates the input features. Various types of features are generated, and the machine learning model is trained. Figure 1. illustrates the flow diagram of proposed method.
It consists of the following steps:
\begin{itemize}
    \item Data Preprocessing
    \item Feature Generation
    \item Constructing ML Model \& Training
    \item Model Serving
\end{itemize}

\begin{table*}[]
\caption{Table 1. Fields of Interest in Bug Report}
\begin{tabular}{ |c||l| }
 \hline
\textbf{Field} & \textbf{Description} \\ \hline
Headline & Concise description of the issue\\ \hline
Description & Detailed outline of the bug \\ \hline
Project & Project \\ \hline
Product & Product \\ \hline
Component & Component \\ \hline
Version & Version \\ \hline
Hardware & Hardware \\ \hline
Severity &  The severity of the report, i.e., high, medium, low, very low\\ \hline
Status & The status of the report, i.e., New, Assigned, Duplicate, Resolved, Closed\\ \hline
\end{tabular}
\end{table*}

Each step of Advaita is explained in the following subsections. 

\subsection{Data Preprocessing}
\vspace{1em}
The headline and detailed description are included in the textual information.
The categorical information contains three parts of the bug reports: product, component and version. The duplication information(status=D) is used to validate the accuracy of the experimental results. A bug report consists of multiple fields. The fields in different projects may vary to some extent, but in general they are similar. Table 1 lists the main fields of a bug report.\par

Since there are some invalid bug reports in the training sample space, they need to be removed. If the headline and description of the bug report are all null, it should be eliminated. The data preprocessing is followed by extracting data including data cleaning, word splitting, stemming, synonyms transforming, stop words filtering, and lower case transforming. \par

Note that, the data cleaning process is different from common text processing, because majority of the bug description contains invalid information expressed in a fixed sentence pattern. Advaita does the following:
\begin{itemize}
    \item lowercase the letters
    \item stop words are removed
    \item Non-Ascii chars are removed.
    \item IPv4/IPv6/MAC address are replaced with “address” keyword.
    \item File path are replaced using “file path” keyword.
    \item Bugs with descriptions of length $<$ 20 are removed.
\end{itemize}

\subsection{Feature Generation}
\vspace{1em}
In Advaita, different features are considered as part of machine learning model. viz.
\newline
\subsubsection{Basic Text Statistical Features}
These features are considered to be classic text features. Using bug text(headline + description) following features are generated.
\begin{itemize}
    \item difference of lengths of two bugs
    \item difference of word counts of two bugs
    \item difference of unique words of two bugs
    \item number of sentences of two bugs
    \item syllable counts of two bugs
    \item char counts of two bugs
    \item common word count of two bugs
    \item Levenshtein distance of two bugs
    \item skew of two bugs
    \item kurtosis of two bugs
\end{itemize}
These features take care of statistical properties of text. Using these features alone Advaita has achieved an accuracy of 70\%. \newline

\subsubsection{Semantic Features}
Counts of noun phrases and verb phrases are taken of each bug text. These counts and their differences highlight the actual meaning of two texts. Adding enhanced its accuracy to about accuracy of 76\%.  \href{https://chartbeat-labs.github.io/textacy/getting_started/quickstart.html}{Textacy} library was used to get the counts.\newline

\subsubsection{Contextual Features}
It uses word embeddings to get the contextual what this means is that “look for” and “find” will annotate to the same meaning. Advaita has gained 90\% accuracy by adding these features. Various distances have been taken as feature vectors viz.

\begin{itemize}
    \item euclidean distance
    \item canberra distances
    \item jaccard distance
    \item cityblock distance
    \item cosine distance
    \item minkowski distance
    \item braycurtis distance
\end{itemize}

Tree based method is used to compute feature importance which ranges from 0 to 1. Figure 2 illustrates feature importance of input features.

\begin{figure*}[ht!] 
\centering
\includegraphics[width=4in]{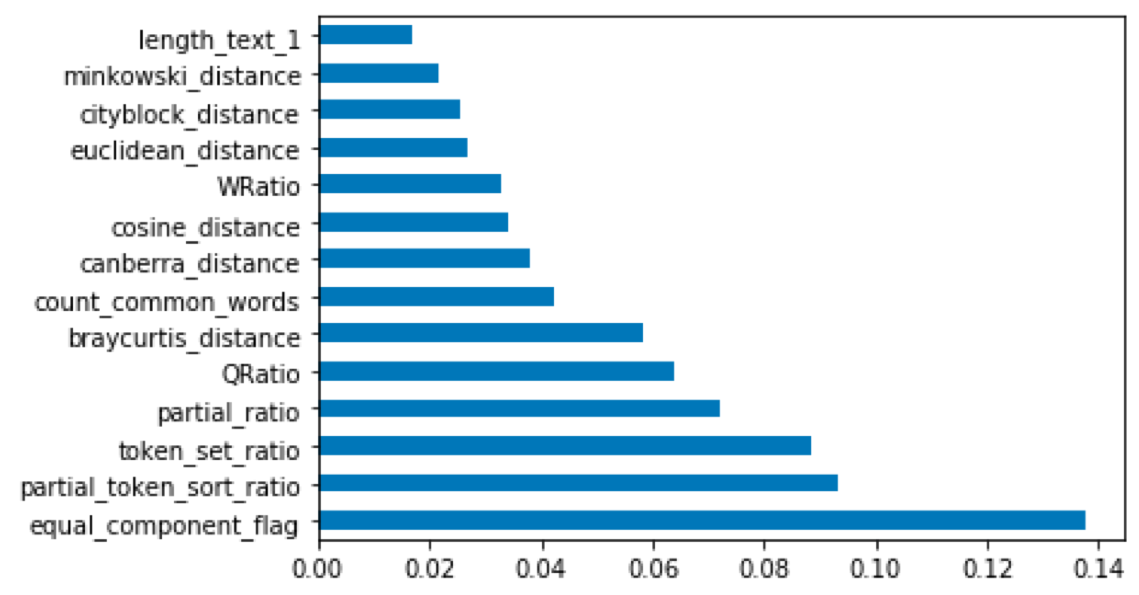}
\caption{Fig 2. Input Feature Importance}
\label{picture}
\end{figure*}

\subsection{Constructing ML Model}
\vspace{1em}

In this Section, application of classification for deciding whether a pair of bug reports are duplicates or not is discussed. The idea is to use machine learning to amplify the impact of the work of the developer; as the developer identifies duplicate bugs, the classifier learns how to better recognize duplicates and may suggest candidates to the developer and thus simplify his/her task.

To retrieve the duplicate bug reports, a well-known machine learning classification algorithm XGBoost is used for performing classification. In each experiment, a table including pairs of bug reports with a set of features (i.e., textual, semantic and contextual features) is passed to the classifiers. Each “all-features” table includes all the inputs necessary for the classifiers to classify each pair of bug reports as duplicate or not. 

\href{https://xgboost.readthedocs.io/en/latest/}XGBoost is used to construct the ML model. To search optimum parameters for XGBoost, grid search from sklearn is applied. All text from training data is converted into feature space and XGBoost model training was done. Log loss was used as loss function due to the binary classification problem.

\begin{equation}
LogLoss = - \frac{1}{n} \sum\limits_{i=1}^n [y_i \cdot log_e(\hat{y_i}) + (1-y_i) \cdot log_e(1-\hat{y_i}) ] 
\end{equation}
\par
where, $y_i$ is the actual label, $\hat{y_i}$ is the predicted label and \textbf{n} is the number of samples in training data.
\newline
Hyper parameters used for XGBoost are:

\begin{table}[H]
\begin{tabular}{|c||c|}
\hline

\textbf{Hyper Parameter} & \textbf{Value} \\ \hline
sub\_sample & 0.8\\ \hline
n\_estimators & 300 \\ \hline
max\_child\_weight & 5 \\ \hline
max\_depth & 4\\ \hline
gamma & 1.5\\ \hline
colsample\_bytree & 0.8 \\ \hline

\end{tabular}
\caption{Table 2. Hyper Parameters of XGBoost}
\end{table}
\subsection{Model Serving}
\vspace{1em}
For a new bug, data preprocessing is done. A dataset is created with new bug as against all the other bugs for a product \& component. Feature generation is applied on this dataset and these features are sent to the deployed model for prediction.\par

Once model is deployed model api is called from app server. Predictions are generated as bug/non-bug probabilities along with bug's data. Using this data user makes a decision to create or use existing resolution from the suggested probable duplicate bugs.

There are three servers deployed for model serving pipeline.
\begin{itemize}
    \item \textbf{App Server}: App server takes requests from front-end and passes them to Model server in json format. The json object includes headline, description, product, project and component. After getting response from Model server App server returns the results to front-end.
    
    \item \textbf{Model Server}: This server provides model api which is an endpoint to serve trained XGBoost model. Model api is called from App server and Model server sends data to trained XGBoost model. Generated results are returned in form of probable duplicate bugs. \par
    
    In terms of model serving scalability, it can be scaled using Clipper, Ray Serve or any other similar frameworks. 
    
    \item \textbf{Embedding Server}: In feature generation step few of the input features are vector distances which are computed using embedding vectors. Bug data (headline + description) is used to generate embeddings. Advaita uses \href{https://fasttext.cc/}{fasttext} to generate embedding. From fasttext 300 dimensional vector is generated. \par 
    
    To facilitate embedding functionality Advaita houses an embedding server. The job of embedding server is to provide embeddings of text on demand. This decoupling of embedding module helps saving time when app or model server is restarted, there is no need to restart embedding server. This saves time as embedding models are large in size and take time to load in memory. Embedding server gives freedom to load multiple embeddings which can be chosen in runtime.\par
    
    In Advaita, Infersentence and Word2Vec embeddings are also explored. Infersentence gives 4000 dimensional vector, because of its large size storage and computing vector distance becomes a challenge. Word2Vec gives 300 dimensional vector but the text embeddings don't help to improve the accuracy. However fasttext embeddings proved to be more efficient in terms of computation and storage. It is easier to modify and plugin any new embeddings like BERT, XLnet in Embedding Server which may improve the accuracy.
\end{itemize}
\vspace{1em}
\section{Results}
\vspace{1em}
In this Section, two use cases are discussed related to the general task of duplicate-bug detection. The first use case refers to a basic decision problem, of whether two specific bug reports are duplicates of each other (given their similarity measurements). The second scenario reflects the general de-duplication scenario, where the incoming bug is compared against all bugs in the repository for a product and component and a ranked list of candidate duplicates is presented to the user who can make the final decision about the real duplicates. Clearly the two scenarios are closely related, since at their core they both assess how similar two bug are. Classification methods, deciding whether a bug is a duplicate of another bug or not, can be combined with ranked prediction, potentially to trim the ranked list.\newline

\subsection{Data Sets}
\vspace{1em}
\subsubsection{Bugs from Cisco Projects} In this experiment bugs from few Cisco projects have been taken. For evaluation 170,000 bugs are taken. After removing invalid bugs, there are 156,000 bugs remained, among which 80,000 bugs are marked as "duplicate". A small test sample(20\% of dataset) is constructed to validate the results.
Table III. describes result metrics obtained from Advaita.

\begin{table}[H]
\begin{tabular}{|c||c|}
\hline

\textbf{Metric} & \textbf{Value} \\ \hline
Precision & 90\% \\ \hline
Recall & 98\% \\ \hline
F1-Score & 94\% \\ \hline
Accuracy & 87\% \\ \hline

\end{tabular}
\caption{Table 3. Result Metrics of Advaita}
\end{table}

\subsubsection{Bugs from Mozilla Firefox Project}
Bugzilla is a bug-tracking system for Firefox. A dataset with over 37000 reported bugs extracted using Bugzilla native REST API \cite{nine} and is used to evaluate Advaita. After removing invalid bugs, there are 36,599 bugs remained, among which 18,000 bugs are marked as "duplicate". A small test sample (20\% of dataset) is constructed to validate the results.
Table IV. describes result metrics obtained from Advaita for Firefox project.

In the Firefox bug repository, both Bug-279438 and Bug-260367 (from test dataset) are
about deleting several history items using ctrl+click or shift+click. Bug-279438 had been
identified as a duplicate of Bug-260367. Specifically, their summaries are as follows. 
\newline

\begin{table}[H]
\begin{tabular}{|c||c|}
\hline

\textbf{Metric} & \textbf{Value} \\ \hline
Precision & 87\% \\ \hline
Recall & 86\% \\ \hline
F1-Score & 87\% \\ \hline
Accuracy & 87\% \\ \hline

\end{tabular}
\caption{Table 4. Result Metrics of Advaita for Firefox bugs}
\end{table}

\textbf{Bug-279438}: This occurs when history is viewed by site last visited. I am not able to
ctrl+click or shift+click to highlight a large chunk of history easily for quick
delete. I was able to do this in 0.9.3 and earlier, but it has stopped working
in 1.0 \newline \par
\textbf{Bug-239223}: I am no longer able to select (using shift-click or control-click) multiple
items in my history or bookmarks to delete them.\newline

Advaita detects these two bugs as duplicate. These two bugs are using different language to convey the same thing. For example pairs like \textit{"highlight"} and \textit{"select"}, \textit{"large chunk"} and \textit{"multiple items"} share the same meaning in this bug context. This context similarity is calculated in context features generation under Advaita.\newline

\subsection{Evaluation Metrics}
\vspace{1em}
To evaluate the performance of the classification to classify bug-report
pairs as duplicates or not, the following metrics are used: accuracy, recall, precision and F1-Score.\\
Precision talks about how precise/accurate your model is. That is, out of those predicted positive, how many of them are actual positives.

\begin{equation}
    \text{Precision}  = \frac{TP}{TP + FP}
\end{equation}
\newline
where, 
$\text{TP} = \text{Duplicate bugs correctly identified}$ and \newline
$\text{FP} = \text{Bugs incorrectly labeled as Duplicates}$

Recall calculates how many of the Actual Positives the model captured through labeling it as a Positive (True Positive).

\begin{equation}
    \text{Recall} = \frac{TP}{TP + FN}
\end{equation}
\newline
where, 
$\text{FN} = \text{Bugs incorrectly labeled as Non-Duplicates}$
\newline

F1 Score is the weighted average of Precision and Recall. Therefore, this score takes both false positives and false negatives into account.

\begin{equation}
    \text{F1-Score} = 2*\frac{\textit{Precision * Recall}}{\textit{precision + Recall}}
\end{equation}
\newline
Accuracy is the proportion of true results (truly classified “dup”s and “non”s) among all pairs being classified. The formula for accuracy is indicated below

\begin{equation}
    \text{Accuracy} = \frac{TP+TN}{TP + FP + TN + FN}
\end{equation}
where, 
$\text{TN} = \text{Non-Duplicate Bugs correctly identified}$\newline

\section{Conclusion}
\vspace{1em}
In this paper, a system is proposed that automatically classifies duplicate bug reports. Advaita considers three factors viz. basic, semantic and contextual text features. These features are fed to XGBoost model. This model well performed on duplicate bug reports written with same textual tokens as well as written with semantically similar ones. This method is empirically evaluated using a dataset of 170,000 bug reports from few Cisco products and 37,000 bugs from Mozilla Firefox project. The results shows that this method performs well and can be implemented in a production environment with little additional effort and a possible non-trivial payoff.  

\bibliographystyle{IEEEtran}
\vspace{1em}
\bibliography{IEEEabrv,vedar}

\begin{thebibliography}{1}
\providecommand{\url}[1]{#1}
\csname url@samestyle\endcsname
\providecommand{\newblock}{\relax}
\providecommand{\bibinfo}[2]{#2}
\providecommand{\BIBentrySTDinterwordspacing}{\spaceskip=0pt\relax}
\providecommand{\BIBentryALTinterwordstretchfactor}{4}
\providecommand{\BIBentryALTinterwordspacing}{\spaceskip=\fontdimen2\font plus
\BIBentryALTinterwordstretchfactor\fontdimen3\font minus
  \fontdimen4\font\relax}
\providecommand{\BIBforeignlanguage}[2]{{%
\expandafter\ifx\csname l@#1\endcsname\relax
\typeout{** WARNING: IEEEtran.bst: No hyphenation pattern has been}%
\typeout{** loaded for the language `#1'. Using the pattern for}%
\typeout{** the default language instead.}%
\else
\language=\csname l@#1\endcsname
\fi
#2}}
\providecommand{\BIBdecl}{\relax}
\BIBdecl

\bibitem{one}
\BIBentryALTinterwordspacing
D.~Cubranic and G.~C. Murphy, ``{Automatic bug triage using text
  categorization},'' 2004. [Online]. Available:
  \url{https://pdfs.semanticscholar.org/169b/b50e92c1a5feb756295e4a9e5492e79cf382.pdf?_ga=2.42512613.1173504973.1579185845-822082848.1578914862}
\BIBentrySTDinterwordspacing

\bibitem{two}
\BIBentryALTinterwordspacing
J.~Anvik, L.~Hiew, and G.~C. Murphy, ``{Who Should Fix This Bug?}'' no.~10, p.
  361–370, 2006. [Online]. Available:
  \url{https://doi.org/10.1145/1134285.1134336}
\BIBentrySTDinterwordspacing

\bibitem{three}
\BIBentryALTinterwordspacing
P.~Bhattacharya and I.~Neamtiu, ``{Fine-grained incremental learning and
  multi-feature tossing graphs to improve bug triaging},'' pp. 1--10, 2010.
  [Online]. Available: \url{https://doi.org/10.1109/ICSM.2010.5609736}
\BIBentrySTDinterwordspacing

\bibitem{four}
\BIBentryALTinterwordspacing
J.~A.-K. T. N.~N. Ahmed~Tamrawi, Tung Thanh~Nguyen, ``{Fuzzy Set and
  Cache-based Approach for Bug Triaging},'' 2011. [Online]. Available:
  \url{http://home.engineering.iastate.edu/~atamrawi/Bugzie/Bugzie.html}
\BIBentrySTDinterwordspacing

\bibitem{five}
\BIBentryALTinterwordspacing
Z.~Lin, F.~Shu, Y.~Yang, C.~Hu, and Q.~Wang, ``{An Empirical Study on Bug
  Assignment Automation Using Chinese Bug Data},'' vol.~1, no.~5, p. 451–455,
  2009. [Online]. Available: \url{https://doi.org/10.1109/ESEM.2009.5315994}
\BIBentrySTDinterwordspacing

\bibitem{six}
\BIBentryALTinterwordspacing
P.~Hooimeijer and W.~Weimer, ``{Modeling Bug Report Quality},'' no.~10, pp.
  34--43, 2007. [Online]. Available:
  \url{https://doi.org/10.1145/1321631.1321639}
\BIBentrySTDinterwordspacing

\bibitem{seven}
\BIBentryALTinterwordspacing
N.~B. S. J. A.~S. Thomas~Zimmermann, Rahul~Premraj and C.~Weiss, ``{What Makes
  a Good Bug Report?}'' 2016. [Online]. Available:
  \url{https://doi.org/10.1109/TSE.2010.63}
\BIBentrySTDinterwordspacing

\bibitem{eight}
\BIBentryALTinterwordspacing
T.~Menzies and A.~Marcus, ``{Automated severity assessment of software defect
  reports},'' pp. 346--355, 2008. [Online]. Available:
  \url{https://doi.org/10.1109/ICSM.2008.4658083}
\BIBentrySTDinterwordspacing

\bibitem{nine}
\BIBentryALTinterwordspacing
``{Bugzilla:REST API}.'' [Online]. Available:
  \url{https://bugzilla-dev.allizom.org/rest/bug/}
\BIBentrySTDinterwordspacing

\end{thebibliography}

\end{document}